\documentclass[conference,10pt]{IEEEtran}
\usepackage{amsmath,amssymb,amsfonts}
\usepackage{algorithmic}
\usepackage{graphicx}
\usepackage{textcomp}
\usepackage{xcolor}
\def\BibTeX{{\rm B\kern-.05em{\sc i\kern-.025em b}\kern-.08em
    T\kern-.1667em\lower.7ex\hbox{E}\kern-.125emX}}

\usepackage{makecell}
\usepackage{booktabs}
\usepackage{xurl} 
\usepackage[backend=biber,
            style=numeric-comp,
            natbib=true,
            doi=false,
            isbn=false,
            url=false,
            date=year,
            maxcitenames=1,
            giveninits=true, 
            sorting=none,
            bibencoding=utf8]{biblatex}
\AtEveryBibitem{\clearfield{note}}
\usepackage{listings}
\usepackage{adjustbox}
\makeatletter
\def\lst@makecaption{\def\@captype{table}\@makecaption}
 
\renewcommand{\fnum@lstlisting}{}
\makeatother

\makeatletter

\usepackage{eso-pic}
\newcommand{\conf}[1]{
\AddToShipoutPictureBG*{
\AtPageUpperLeft{
 \put(\LenToUnit{0pt},\LenToUnit{-1cm}){
     \parbox{\paperwidth}{\centering\fontsize{9}{11}\selectfont\itshape #1}}
 }}}
\newcommand{\notice}[1]{
\AddToShipoutPictureBG*{
\AtPageLowerLeft{
 \put(\LenToUnit{\oddsidemargin + 1in},\LenToUnit{1cm}){
     \parbox{\textwidth}{\centering\fontsize{7}{9}\selectfont #1}}
 }}}
\conf{This work has been accepted for the 2025 Picture Coding Symposium (PCS).\\The final published version will be available via IEEE Xplore.}
\notice{\copyright{} 2025 IEEE. Personal use of this material is permitted. Permission from IEEE must be obtained for all other uses, in any current or future media, including reprinting/republishing this material for advertising or promotional purposes, creating new collective works, for resale or redistribution to servers or lists, or reuse of any copyrighted component of this work in other works}

\usepackage[pdftex,
            pdfauthor={Benjamin Herb, Rakesh Rao Ramachandra Rao, Steve Göring and Alexander Raake},
            pdftitle={Evaluating Video Quality Metrics for Neural and Traditional Codecs using 4K/UHD-1 Videos}
            ]{hyperref}

\begin{document}

\title{Evaluating Video Quality Metrics for Neural and Traditional Codecs using 4K/UHD-1 Videos}

\author{%
	\IEEEauthorblockN{{Benjamin Herb\IEEEauthorrefmark{1}, Rakesh Rao Ramachandra Rao\IEEEauthorrefmark{1}, Steve G\"oring\IEEEauthorrefmark{1}, Alexander Raake\IEEEauthorrefmark{2}} \\
		\IEEEauthorblockA{{\IEEEauthorrefmark{1}Audiovisual Technology Group, Technische Universität Ilmenau, Germany}}
		\IEEEauthorblockA{{\IEEEauthorrefmark{2}Institute for Communications Engineering (IENT), RWTH Aachen, Germany}}
		}
		{Email: [benjamin.herb, rakesh-rao.ramachandra-rao, steve.goering]@tu-ilmenau.de, raake@ient.rwth-aachen.de}
	}

\maketitle

\begin{abstract}
With neural video codecs (NVCs) emerging as promising alternatives for traditional compression methods, it is increasingly important to determine whether existing quality metrics remain valid for evaluating their performance.
However, few studies have systematically investigated this using well-designed subjective tests.
To address this gap, this paper presents a subjective quality assessment study using two traditional (AV1 and VVC) and two variants of a neural video codec (DCVC-FM and DCVC-RT).
Six source videos (8-10 seconds each, 4K/UHD-1, 60 fps)  were encoded at four resolutions (360p to 2160p) using nine different QP values, resulting in 216 sequences that were rated in a controlled environment by 30 participants.
These results were used to evaluate a range of full-reference, hybrid, and no-reference quality metrics to assess their applicability to the induced quality degradations.
The objective quality assessment results show that VMAF and AVQBits\textbar H0\textbar f demonstrate strong Pearson correlation, while FasterVQA performed best among the tested no-reference metrics.
Furthermore, PSNR shows the highest Spearman rank order correlation for within-sequence comparisons across the different codecs.
Importantly, no significant performance differences in metric reliability are observed between traditional and neural video codecs across the tested metrics.
The dataset, consisting of source videos,  encoded videos, and both subjective and quality metric scores will be made publicly available following an open-science approach\footnote{\url{https://github.com/Telecommunication-Telemedia-Assessment/AVT-VQDB-UHD-1-NVC}}.
\end{abstract}

\begin{IEEEkeywords}
video quality assessment, deep learning, neural video coding, video quality metrics, subjective evaluation, dataset, DCVC, AV1, VVC, 4K, UHD
\end{IEEEkeywords}

\section{Introduction} 

In recent years deep learning (DL) has been increasingly integrated into various image and video processing tasks, showing significant improvements over conventional algorithmic approaches.
Efficient video coding is one such task, with applications in a variety of fields, including online video streaming.
This is particularly important because video streaming makes up a significant portion of overall internet usage, accounting for 65\% of total internet volume in 2023\footnote{\url{https://www.applogicnetworks.com/press-releases/sandvines-2023-global-internet-phenomena-report-shows-24-jump-in-video-traffic-with-netflix-volume-overtaking-youtube}}.

Traditional codecs, including more recent ones such as AV1 and VVC (H.266), employ conventional hybrid video coding layouts. 
Recently developed deep learning-based codecs, also referred to as neural video codecs (NVCs), have been designed to either entirely replace the conventional codec with network architectures or to substitute specific components.

DeepCoder, one of the first NVCs proposed by \citeauthor{chen_deepcoder_2017}, used a convolutional neural network (CNN) based video compression framework with a fixed block size of 32$\times$32 to achieve comparable performance to H.264 in terms of SSIM~\cite{chen_deepcoder_2017}.
Further, \citeauthor{park_deep_2020}~\cite{park_deep_2020} proposed DeepPVCnet, a NVC with bi-directional prediction, which showed performance comparable to H.264 and H.265 in terms of PSNR and MS-SSIM.
\citeauthor{montajabi_recurrent_2022}~\cite{montajabi_recurrent_2022} proposed a recurrent neural network (RNN) based video codec that outperforms both H.264 and H.265 across metrics such as PSNR, SSIM, and VMAF.
The first generative adversarial network (GAN) based video codec was proposed by \citeauthor{mentzer_neural_2022}~\cite{mentzer_neural_2022}. 
The authors report that typical quality metrics cannot be fully relied on to assess the performance of NVCs and proposed user studies and the development of perceptual metrics that take the ``newer'' distortions introduced by NVCs into account.
In addition to this, there have been NVCs that are iteratively developed to improve the compression efficiency and also target specific use cases.
Notable examples include the DCVC family of codecs~\cite{li_deep_2021, sheng_temporal_2022, li_hybrid_2022, li_neural_2023, li_neural_2024, jia_towards_2025, wang_evc_2023} and DHVC~\cite{lu_deep_2024, lu_high-efficiency_2024}, both of which consistently perform either on-par with or outperform traditional video codecs such as H.265, at least in terms of PSNR.

Several studies have also evaluated variants of DCVC for different applications.
\citeauthor{teng_benchmarking_2024} \cite{teng_benchmarking_2024} compared neural (DCVC-FM, DCVC-DC) and traditional compression methods (AV1, VVC, AVM, ECM) configured for low delay applications using VMAF and PSNR.
\citeauthor{regensky_analysis_2024} \cite{regensky_analysis_2024} compared the compression performance of four DCVC variants (DCVC, DCVC-TCM, DCVC-HEM, and DCVC-DC) to VVC for 360-degree videos.

For most of the codec development and the comparative studies the performance has been assessed only in terms of objective metrics such as PSNR, SSIM, MS-SSIM, and VMAF.
However, this may not reflect the efficacy of the codecs in terms of subjective quality, as NVCs may introduce new types of distortions.
This unreliability of metrics was also highlighted by \citeauthor{mentzer_neural_2022}~\cite{mentzer_neural_2022}.

In this paper, we conduct a visual quality assessment study to compare the impact of distortions from deep learning-based and traditional video codecs, to better understand whether the new distortions introduced by NVCs affect the perceived video quality.
The results will be used to assess prediction performance of a number of objective metrics, such as VMAF, and evaluate the need to adapt them for this new context.

\section{Subjective Assessment}
\subsection{Source Videos}
\begin{figure}
  \centering
  \includegraphics[width=1\linewidth]{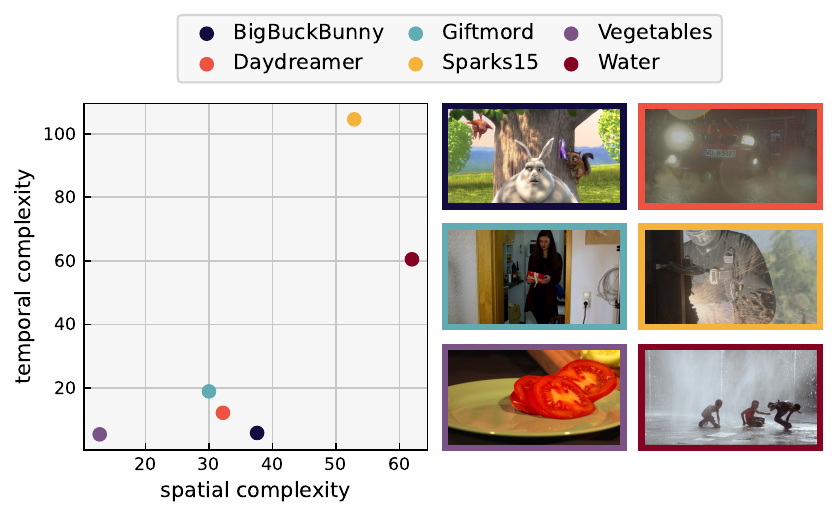}
  \caption{Mean spatial and temporal complexity of the selected videos calculated using VCA \cite{menon_vca_2022} on the left with stills from the videos on the right.}
  \label{fig:vca}
\end{figure}

Six 8-10 second video clips were selected from the AVT-VQDB-UHD-1 \cite{ramachandra_rao_avt-vqdb-uhd-1_2019} dataset with a resolution of 3840$\times$2160 (4K/UHD-1), YUV420p 8-bit pixel format and a framerate of 60 fps.
Figure \ref{fig:vca} shows the complexity analysis results obtained using the Visual Complexity Analyzer (VCA) \cite{menon_vca_2022}.
The videos were chosen to cover a variety of complexity levels. 

\subsection{Encoding Setup and Configuration}

For this test, four encoders were used: AOMedia Project AV1 Encoder v3.12.0 for AV1, vvencFFapp \cite{wieckowski_vvenc_2021} v1.13.1 for VVC, DCVC-FM \cite{li_neural_2024} (Commit: b67129d), and DCVC-RT \cite{jia_towards_2025} (Commit: 9b7acf7).

To determine practical encoding parameters, several tests were conducted across all source videos at different quality levels and 1080p.
Since both DCVC-FM and DCVC-RT do not provide an option to automatically determine intra frame locations, intra periods of 32, 64, 96, and -1 (one I-frame at the start) were tested.
As expected, shorter intra periods generally result in higher bitrates, with a Bjøntegaard Delta Rate (BD-Rate) \cite{bjontegaard_calculation_2001} increase of up to 20\% when going from -1 to 32.
Given the longer test sequences (480 - 600 frames), an intra period of 96 (instead of the typical -1) was selected for the subsequent encodings as a practical compromise, resulting in a BD-Rate below 6\% for all codecs.
The traditional encoders offer different speed settings that were tested as well.
For practical implementation considerations, VVenC was configured to use the medium profile (BD-Rate 6.6\% / 5.7$\times$ Speed compared to the slowest option), while AOM's \textit{cpu-used} parameter was set to 4 (BD-Rate 10.72\% / 22.1$\times$ Speed). 
Both codecs use random access (hierarchical) configurations (AOM Common Test Conditions \cite{alliance_for_open_media_aom_2022} for AV1 and \textit{randomAccess.cfg} for VVenC) to ensure that the testing conditions represent realistic usage scenarios.
DCVC-FM and DCVC-RT only offer low-delay P inter-frame configurations, which inherently constrains their compression efficiency compared to hierarchical approaches.
However, this experimental design prioritizes evaluating each codec using practical configurations instead of enforcing uniform constraints.
This type of comparison is also recommended by the developers of DCVC \cite{microsoft_test_2023}.

\subsection{Test Implementation}

\begin{figure}[h]
  \centering
  \includegraphics[width=0.9\linewidth]{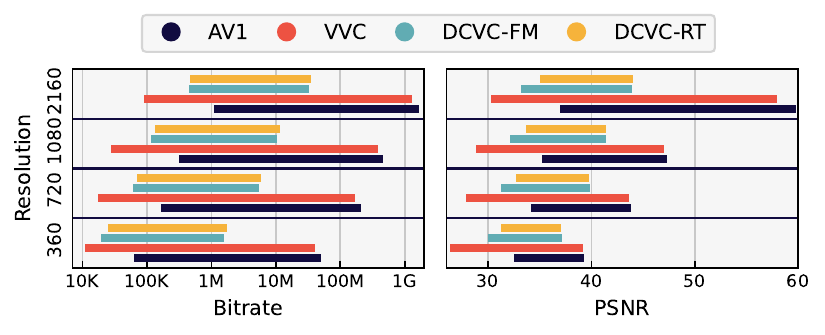}
  \caption{Maximum and minimum bitrate / PSNR ranges for each codec and resolution when using the entire range of quality parameters [0-63].}
  \label{fig:codec_ranges}
\end{figure}

All codecs provide quality levels from 0 - 63, however, for DCVC-FM and DCVC-RT higher values correspond to higher quality, which is reversed from the traditional codecs.
Figure \ref{fig:codec_ranges} illustrates the maximum achievable bitrate and quality ranges for each codec across different resolutions when encoding the six sequences.

\begin{figure*}
  \centering
  \includegraphics[width=\linewidth]{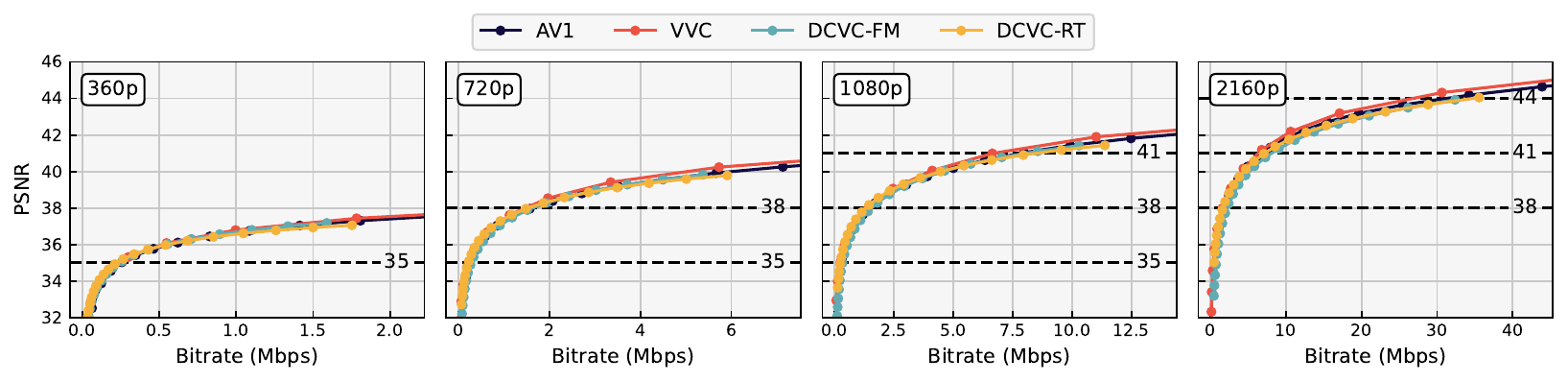}
  \caption{Mean PSNR and bitrate values for all six sequences encoded at multiple quality levels per codec and resolution. Quality levels are chosen using the target PSNR values and interpolating the closest level for each codec.}
  
  \label{fig:quality_selection}
\end{figure*}

A quality-based selection method was implemented to select the appropriate encoding parameters for each codec.
The source videos were encoded at 21 quality levels for each resolution, the result of which can be seen in Figure \ref{fig:quality_selection}.
Three target PSNR values were selected, covering a wide quality range for 2160p and 1080p, with ranges limited by the highest achievable quality of DCVC-FM and DCVC-RT and the lowest quality of AV1.
For 720p and 360p, fewer parameters were chosen to avoid overloading the test.
Based on the target PSNR values, the closest QP values were interpolated.
The selected quality parameters are documented in Table \ref{tab:qualities}, which were applied to all source videos, resulting in 216 processed video sequences (PVS).
Compared to a bitrate-based selection approach, this ensures similar quality ranges across all test sequences.
However, this also inherently results in substantial bitrate variations between the different source videos.

\begin{table}[h]
\caption{Selected quality parameters for each codec and resolution based on mean PSNR targets.}
\centering
  \label{tab:qualities}
  \adjustbox{width=0.8\linewidth,center}{
  \begin{tabular}{l|c|cc|ccc|ccc}
    \toprule
    \multicolumn{1}{c}{} & \multicolumn{1}{c}{360p} & \multicolumn{2}{c}{720p} & \multicolumn{3}{c}{1080p} & \multicolumn{3}{c}{2160p} \\
    \midrule
    \textbf{PSNR} [dB] & 35 & 38 & 35 & 41 & 38 & 35 & 44 & 41 & 38 \\
    \midrule
    \textbf{AV1}      & 54   & 48 & 61       & 36 & 55 & 63       & 31 & 50 & 61       \\
    \textbf{VVC}      & 34   & 32 & 41       & 27 & 36 & 45       & 25 & 34 & 42       \\
    \textbf{DCVC-FM}  & 38   & 46 & 25       & 59 & 37 & 18       & 63 & 43 & 26       \\
    \textbf{DCVC-RT}  & 34   & 42 & 17       & 58 & 32 & 10       & 63 & 39 & 19       \\
    \bottomrule
  \end{tabular}
  }
\end{table}

Due to a technical error, eight PVS of \textit{Sparks15} in 720p were encoded with 280 instead of 480 frames (4.5s instead of 8s).
Comparing the short and regular versions resulted in a mean absolute PSNR difference of 0.30 dB  (max. 0.35 dB) and 3.36 for VMAF (max. 4.64).
These eight short versions were used for subsequent testing, as they were the versions shown in the subjective evaluation.

\begin{figure}
  \centering
  \includegraphics[width=\linewidth]{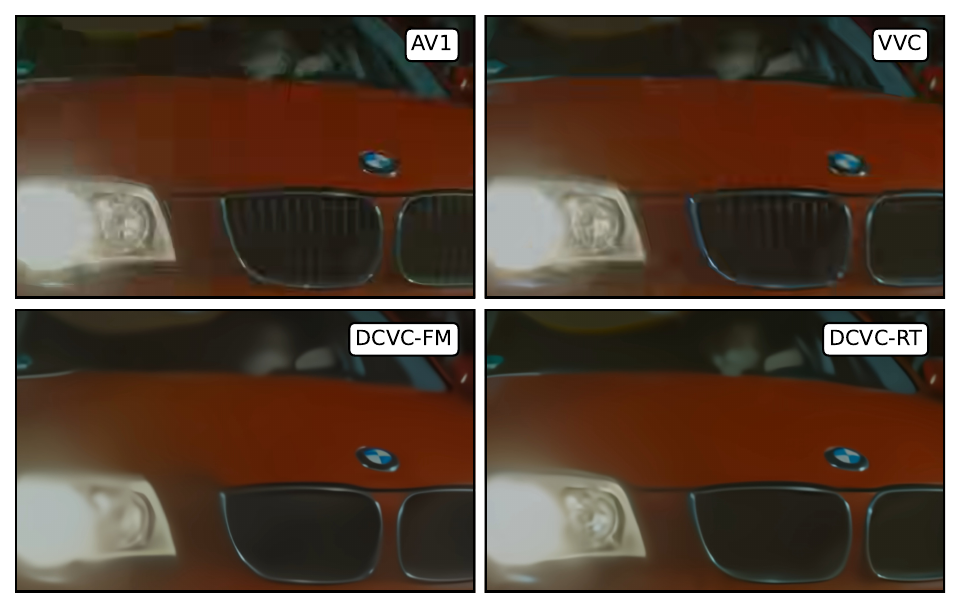}
  \caption{Example crops from \textit{Daydreamer} at the lowest quality setting (360p).}
  \label{fig:example_encoded}
\end{figure}

Figure \ref{fig:example_encoded} shows exemplary results from the lowest quality setting (360p) for \textit{Daydreamer}. 
Notable are the visible blocking artifacts with AV1 and VVC, while the DCVCs produce smoother images.
Some details, like the car's logo, are better preserved by DCVC, while others, such as the grille, are preserved by AV1 and VVC but lost using DCVC.

\subsection{Experimental Procedure}

The test was conducted in a controlled environment on an Asus XG43UQ UHD Monitor (43 inch), with a fixed viewing distance of 1.5H.
Ratings were collected with avrateNG\footnote{\url{https://github.com/Telecommunication-Telemedia-Assessment/avrateNG}} using mpv\footnote{\url{https://mpv.io/}} for playback. 
Each video was rated using the 5-point absolute category rating (ACR) \cite{itu-t_p910_2023} method with testing lasting 45 minutes per participant.
Before testing, each participant completed a FrACT10 vision test\footnote{\url{https://michaelbach.de/fract/}}.

The study was conducted on 30 paid participants (students and employees of the university).
Each participant rated all 216 PVS, presented in a random order.
However, due to a technical issue, 33 of the 6480 total ratings were not captured correctly and subsequently removed from the dataset. 
This resulted in each individual PVS having between 28 and 30 ratings.
To ensure the reliability of the participants, an outlier detection according to ITU-T P.910 \cite{itu-t_p910_2023} was applied.
The Pearson correlation of each subject's results and the mean opinion score (MOS) were calculated.
Participants with a $PCC < 0.75$ were discarded, starting with the lowest one and recalculating the MOS after each removal.
All further analysis is based on the 26 participants who passed this outlier detection.

\subsection{Subjective Results}

\begin{figure}
  \centering
  \begin{minipage}[b]{0.49\linewidth}
    \centering
    \includegraphics[width=1\linewidth]{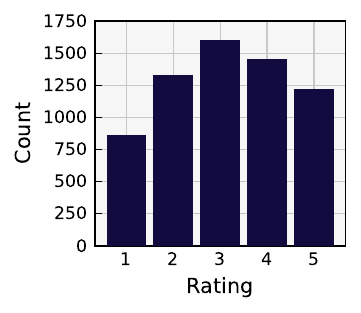}
    \caption{Distribution of ratings.}
    \label{fig:rating_distribution}
  \end{minipage}
  \hfill
  \begin{minipage}[b]{0.49\linewidth}
    \centering
    \includegraphics[width=1\linewidth]{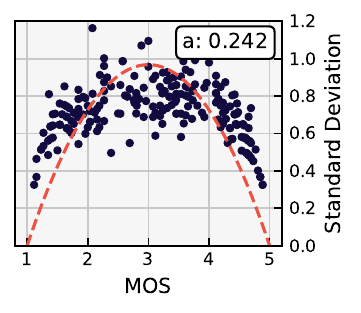}
    \caption{SOS \cite{hossfeld_sos_2011} Analysis.}
    \label{fig:sos}
  \end{minipage}
\end{figure}

\begin{figure}
  \centering
  \includegraphics[width=1\linewidth]{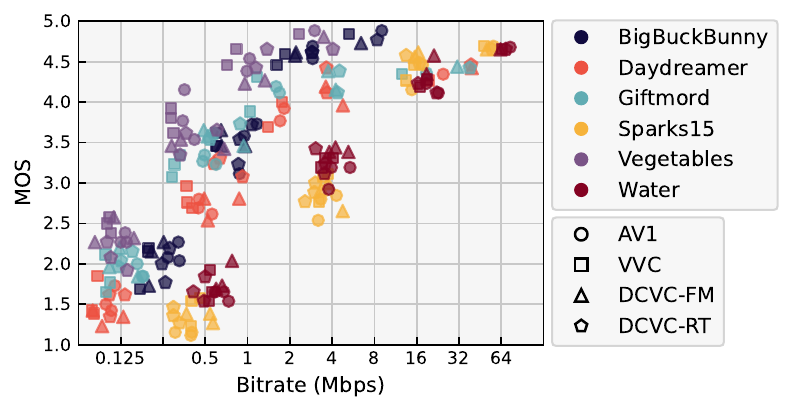}
  \caption{Subjective results for all shown sequences.}
  \label{fig:bitrate_vs_mos}
\end{figure}

The resulting rating distribution can be seen in Figure~\ref{fig:rating_distribution}, which shows an approximately normal distribution.
Additionally, a standard deviation of opinion scores (SOS) \cite{hossfeld_sos_2011} analysis was done on the data, with results shown in Figure \ref{fig:sos}.
The resulting $a$ of 0.242 is comparable to similar studies~\cite{rao_large-scale_2025}.
The overall subjective quality results are shown in Figure \ref{fig:bitrate_vs_mos}.
The different bitrate requirements for different sources are clearly observable, with \textit{Water} and \textit{Sparks15} demonstrating substantially higher bitrate demands.
This aligns with expectations based on the high temporal complexity indicated in the VCA analysis.
Conversely, \textit{Vegetables}, the source video with the lowest complexity, has only one PVS below a MOS of 2.

\section{Objective Quality Assessment}

Current video quality metrics are primarily designed and optimized for predicting the perceptual quality of videos encoded using traditional codecs such as AV1 and VVC.
The following section evaluates different full-reference (FR), no-reference (NR), and hybrid models to assess their applicability to NVCs.

\begin{figure*}[t]
  \centering
  \includegraphics[width=\linewidth]{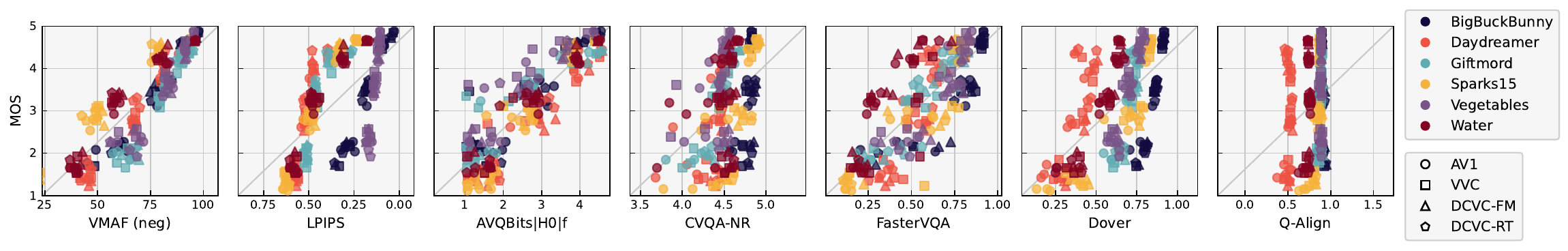}
  \caption{MOS compared to the metric results. Each metric axis is linearly mapped to the ACR scale following ITU-T
Rec. P.1401 \cite{itu-t_p1401_2020}.}
  \label{fig:metric_correlation_and_corr}
\end{figure*}

They include seven FR metrics (PSNR, SSIM, MS-SSIM, VMAF (including the no enhancement gain (neg) variant), CVQA-FR~\cite{sun_deep_2021}, and LPIPS~\cite{zhang_unreasonable_2018}), five NR metrics (MUSIQ~\cite{ke_musiq_2021}, CVQA-NR~\cite{sun_deep_2021}, FasterVQA~\cite{wu_neighbourhood_2023}, Dover~\cite{wu_exploring_2023} and Q-Align~\cite{wu_q-align_2024}), and one hybrid model (AVQBits\textbar H0\textbar f~\cite{ramachandra_rao_avqbitsadaptive_2022}), which uses encoding metadata (bitrate, resolution, and framerate) in addition to the PVS.
The results are compared to the MOS using Pearson correlation coefficient (PCC), Spearman rank order correlation coefficient (SRCC), as well as root mean square error (RMSE) in Table \ref{tab:codec_correlations} and visualized in Figure \ref{fig:metric_correlation_and_corr}.
For RMSE, the values of each metric are linearly mapped to the 5-point ACR scale according to ITU-T Rec. P.1401 \cite{itu-t_p1401_2020}.

\setlength{\tabcolsep}{3pt} 
\begin{table*}[t]
\caption{Correlation between MOS and metric for each codec, the mean correlation (within-sequence) for each source and correlation across all videos. $\Delta NvT$ quantifies the degree to which quality metrics overestimate neural compared to traditional codec performance across equivalent quality levels.}
\tiny
\centering
\label{tab:codec_correlations}
\adjustbox{width=1\linewidth,center}{
\begin{tabular}{lcc||cc||ccc|ccc|ccc|ccc||ccc||ccc}
\toprule
\multicolumn{3}{c}{} & \multicolumn{2}{c}{$\Delta$NvT}& \multicolumn{3}{c}{AV1} & \multicolumn{3}{c}{VVC} & \multicolumn{3}{c}{DCVC-FM} & \multicolumn{3}{c}{DCVC-RT} & \multicolumn{3}{c}{\textbf{Within-Sequence}} & \multicolumn{3}{c}{\textbf{Overall}} \\
  Metric & & & Mean &STD & PCC & SRCC & RMSE & PCC & SRCC & RMSE & PCC & SRCC & RMSE & PCC & SRCC & RMSE & PCC & SRCC & RMSE & PCC & SRCC & RMSE \\
\midrule
PSNR  & FR&Error& -0.055 &0.169 & 0.772 & 0.789 & 0.720 & 0.759 & 0.769 & 0.714 & 0.737 & 0.756 & 0.761 & 0.734 & 0.762 & 0.768 & 0.958 & \textbf{\underline{0.953}} & 0.311 & 0.750 & 0.768 & 0.742 \\
SSIM  & FR&IQA& -0.078 &0.179 & 0.718 & 0.842 & 0.790 & 0.693 & 0.852 & 0.790 & 0.719 & 0.862 & 0.783 & 0.696 & 0.840 & 0.813 & 0.964 & 0.936 & 0.280 & 0.705 & 0.851 & 0.797 \\
MS-SSIM  & FR&IQA& -0.073 &0.178 & 0.714 & 0.776 & 0.794 & 0.688 & 0.784 & 0.796 & 0.695 & 0.752 & 0.810 & 0.686 & 0.776 & 0.823 & \textbf{\underline{0.978}} & 0.937 & \textbf{\underline{0.223}} & 0.695 & 0.774 & 0.808 \\
VMAF & FR&VQA& -0.031 &0.153 & \textbf{0.902} & \textbf{0.919} & \textbf{0.489} & \textbf{0.883} & \textbf{0.902} & \textbf{0.514} & 0.885 & \textbf{0.891} & 0.524 & 0.877 & \textbf{0.906} & 0.544 & 0.970 & 0.940 & 0.251 & 0.886 & \textbf{0.907} & 0.520 \\
VMAF (neg) & FR&VQA& -0.040 &0.152 & \textbf{\underline{0.905}} & \textbf{\underline{0.920}} & \textbf{\underline{0.483}} & \textbf{\underline{0.886}} & \textbf{\underline{0.903}} & \textbf{\underline{0.509}} & \textbf{0.888} & \textbf{\underline{0.894}} & \textbf{0.518} & \textbf{0.880} & \textbf{\underline{0.911}} & \textbf{0.537} & \textbf{0.971} & \textbf{0.942} & \textbf{0.248} & \textbf{\underline{0.889}} & \textbf{\underline{0.909}} & \textbf{\underline{0.514}} \\
LPIPS\cite{zhang_unreasonable_2018} & FR&IQA& -0.069 &0.187 & 0.679 & 0.735 & 0.833 & 0.686 & 0.736 & 0.798 & 0.629 & 0.690 & 0.876 & 0.591 & 0.694 & 0.912 & 0.934 & 0.920 & 0.360 & 0.646 & 0.716 & 0.857 \\
CVQA-FR \cite{sun_deep_2021} & FR&VQA& -0.110 &0.216 & 0.827 & 0.825 & 0.638 & 0.809 & 0.848 & 0.644 & 0.813 & 0.831 & 0.656 & 0.820 & 0.849 & 0.647 & 0.942 & 0.903 & 0.333 & 0.814 & 0.840 & 0.651 \\
AVQBits\textbar H0\textbar f \cite{ramachandra_rao_avqbitsadaptive_2022} & Hybrid&VQA& 0.083 &0.308 & 0.899 & 0.886 & 0.498 & 0.842 & 0.804 & 0.592 & \textbf{\underline{0.909}} & 0.866 & \textbf{\underline{0.470}} & \textbf{\underline{0.924}} & 0.887 & \textbf{\underline{0.433}} & 0.934 & 0.900 & 0.374 & \textbf{0.887} & 0.861 & \textbf{0.518} \\
\midrule
MUSIQ \cite{ke_musiq_2021} & NR&IQA& 0.064 &0.219 & 0.703 & 0.723 & 0.807 & \textbf{0.688} & \textbf{0.726} & \textbf{0.795} & \textbf{0.653} & \textbf{0.667} & \textbf{0.854} & \textbf{0.618} & \textbf{0.623} & \textbf{0.889} & \textbf{\underline{0.921}} & \textbf{\underline{0.896}} & \textbf{\underline{0.416}} & \textbf{0.664} & \textbf{0.683} & \textbf{0.839} \\
CVQA-NR \cite{sun_deep_2021} & NR&VQA& 0.286 &0.411 & \textbf{0.705} & \textbf{0.742} & \textbf{0.805} & 0.460 & 0.489 & 0.974 & 0.416 & 0.462 & 1.025 & 0.344 & 0.365 & 1.062 & 0.550 & 0.641 & 0.872 & 0.469 & 0.491 & 0.992 \\
FasterVQA \cite{wu_neighbourhood_2023} & NR&VQA& -0.006 &0.304 & \textbf{\underline{0.822}} & \textbf{\underline{0.813}} & \textbf{\underline{0.646}} & \textbf{\underline{0.798}} & \textbf{\underline{0.806}} & \textbf{\underline{0.661}} & \textbf{\underline{0.774}} & \textbf{\underline{0.787}} & \textbf{\underline{0.714}} & \textbf{\underline{0.826}} & \textbf{\underline{0.831}} & \textbf{\underline{0.638}} & 0.878 & 0.882 & 0.516 & \textbf{\underline{0.802}} & \textbf{\underline{0.803}} & \textbf{\underline{0.670}} \\
Dover \cite{wu_exploring_2023} & NR&VQA& 0.101 &0.213 & 0.700 & 0.677 & 0.810 & 0.643 & 0.634 & 0.840 & 0.611 & 0.609 & 0.892 & 0.588 & 0.609 & 0.915 & \textbf{0.897} & \textbf{0.884} & \textbf{0.463} & 0.634 & 0.629 & 0.868 \\
Q-Align \cite{wu_q-align_2024} & NR&VQA& -0.004 &0.200 & 0.406 & 0.467 & 1.036 & 0.302 & 0.435 & 1.045 & 0.163 & 0.085 & 1.112 & 0.096 & 0.093 & 1.126 & 0.358 & 0.309 & 0.978 & 0.245 & 0.263 & 1.088 \\
\bottomrule
\end{tabular}
}
\end{table*}

VMAF performs nearly as well on neural video codecs as on traditional codecs, achieving high overall PCC and SRCC around 0.9, with the neg variant performing slightly better.
CVQA-FR also shows good performance, while LPIPS performs worst among the tested FR metrics, especially overestimating low complexity videos (see Fig. \ref{fig:metric_correlation_and_corr}).
SSIM achieves high SRCC, indicating a clear monotonic relationship.
The hybrid model AVQBits\textbar H0\textbar f shows a similarly high PCC as VMAF but lower SRCC, with high variance at lower quality levels, partly due to limitations of the internal HEVC reencode at very low bitrates.
The correlation for NR metrics is lower, with the transformer-based models (FasterVQA, MUSIQ, and the technical branch of Dover) outperforming the rest.
FasterVQA demonstrates the highest overall correlation with PCC and SRCC of around 0.8, while MUSIQ reaches PCC of 0.67.
Dover's technical branch alone achieves PCC of 0.71, but fusing it with the CNN based aesthetic branch (PCC of 0.50) reduces the combined score to 0.63.
CNN based CVQA-NR shows weak correlation for any codec besides AV1, while LLM-based Q-Align shows limited overall performance, likely due to the heavy feature abstraction leading to similar predictions for all PVS from the same source sequences (see Fig. \ref{fig:metric_correlation_and_corr}).
Overall, none of the models show a large performance drop when comparing neural and traditional codecs.

Different codecs, encoders, or parameters are commonly compared using simple metrics such as PSNR on a given video sequence.
Table \ref{tab:codec_correlations} shows the mean of these within-sequence correlations for each metric.
The results demonstrate that computationally less complex tools, like  PSNR, SSIM, and MS-SSIM perform very well even when comparing neural to traditional codecs.
PSNR achieves the highest Spearman correlation in this test, making it a viable choice for comparing different encodings of the same source content.

To identify potential differences in metric performance, the $\Delta NvT$ was calculated as follows:
\begin{align*}
\Delta NvT &= (Metric_{N} - Metric_{T}) - (MOS_{N} - MOS_{T}) 
\end{align*}
where $Metric_{N/T}$ represents the average metric values for neural and traditional codecs for each quality / resolution combination linearly mapped to MOS.
This metric quantifies the degree to which a given metric overestimates ($\Delta NvT>0$) or underestimates the quality of neural compared to traditional codecs.
Most results fall between $\pm$ 0.6 without favoring either codec type, with two exceptions: AVQBits\textbar H0\textbar f shows $\Delta NvT$ between 0.6-1.0 for four low quality 2160p sequences due to the previously mentioned reencoding issues, while CVQA-NR shows $\Delta NvT$ of up to 1.6 at lower qualities, as it predicts similar scores across quality levels for most codecs while more accurately predicting lower quality AV1 scores.
Beyond these outliers, the mean results in Table \ref{tab:codec_correlations} confirm the previous findings that there are no substantial differences in metric estimations between the neural and traditional codecs considered in this study.

\section{Conclusion}
This paper presents a subjective and objective quality evaluation study using two traditional (AV1 \& VVC) as well as two neural video codecs (DCVC-FM \& DCVC-RT) to determine the applicability of different video quality metrics on both codec types.
The full-reference metric VMAF, along with the hybrid model AVQBits\textbar H0\textbar f achieve high PCC of around 0.89 across all sequences and FasterVQA outperforms the other no-reference models with a PCC of 0.8.
Furthermore, PSNR demonstrates the highest within-sequence SRCC result, confirming its utility for evaluating different codecs on a given source sequence.
Notably, the results indicate no significant impact on the performance of the metrics when using the selected neural video codecs compared to traditional ones.
While there remains a clear need for improved no-reference metrics, the study does not reveal any new requirements unique to neural video codecs.
Future work is needed to investigate whether these findings generalize to both a broader range of source sequences and neural video codecs.

\printbibliography

\end{document}